\documentclass[11pt]{article}
\pdfoutput=1
\usepackage{setspace}
\usepackage{graphicx}
\usepackage[space]{grffile}
\usepackage{amsmath}
\usepackage{url}
\usepackage{amsthm}
\usepackage{natbib}
\usepackage{mathtools}
\usepackage{bm}
\usepackage{amssymb}
\usepackage{threeparttable}
\DeclarePairedDelimiter{\abs}{\lvert}{\rvert}
\usepackage{booktabs}
\usepackage{array}
\usepackage[english]{babel}
\usepackage{longtable}
\usepackage{subfig}
\usepackage{ragged2e}
\newtheorem{theorem}{Theorem}[section]

\newtheorem{proposition}[theorem]{Proposition}

\theoremstyle{definition}
\newtheorem{definition}{Definition}[section]

\theoremstyle{remark}

\newcommand{\tabincell}[2]{\begin{tabular}{@{}#1@{}}#2\end{tabular}}
 \usepackage[titletoc]{appendix}
\usepackage{tabularx}
\usepackage{changepage}
\usepackage{verbatim}
\usepackage[a4paper, total={6in, 8in}]{geometry}
\usepackage{setspace}
\usepackage{booktabs}
\usepackage{geometry}
\usepackage{multirow}

\numberwithin{equation}{section}
\usepackage{geometry}
\usepackage{hyperref}
\hypersetup{
	colorlinks,
	linkcolor={blue},
	citecolor={blue},
	urlcolor={blue}
}

\begin{document}

\author{
Zhengxiao Li \footnote{School of Insurance and Economics, University of International Business and Economics, China.}
\quad Jan Beirlant\footnote{Dept. of Mathematics, LStat and LRisk, KU Leuven, Belgium, and Dept. of Mathematical Statistics and Actuarial Science, Univ. of the Free State, South Africa.}
\quad Shengwang Meng\footnote{Corresponding author: School of Statistics, Renmin University, China. E-mail addresses: mengshw@ruc.edu.cn.}
}
\date{}
\title{
Generalizing the log-Moyal distribution and regression models for heavy tailed loss data
}

\maketitle
\begin{abstract}
Catastrophic loss data are known to be heavy-tailed.
Practitioners then need models that are able to capture both tail and modal parts of claim data.
To this purpose, a new parametric family of loss distributions is proposed as a gamma mixture of the generalized log-Moyal distribution from \cite{bhati2018generalized}, termed the generalized log-Moyal gamma distribution (GLMGA). We discuss the probabilistic characteristics of the GLMGA, and statistical estimation of the parameters through maximum likelihood. While the GLMGA distribution is a special case of the GB2 distribution, we show that this simpler model is effective in regression modelling of large and modal loss data.
A fire claim data set reported in \cite{cummins1990applications} and a Chinese earthquake loss data set
are used to illustrate the applicability of the proposed model.
\\
{\bf{JEL Classification Numbers:}}
	
\end{abstract}

\newpage
\section{Introduction}\label{section:introduction}

Modelling of extreme risk data is of  great interest to actuaries in order to calculate risk measures such as Value-at-Risk (VaR) and conditional tail expectation (CTE), or to
determine premiums and reserves and optimal retention
levels in reinsurance and catastrophe insurance. While univariate loss models are well developed, the use of covariate information in order to predict heavy tailed loss data through  regression models has received much less attention, next to traditional generalized linear modelling  such as gamma  and inverse Gaussian regression without specific interest in heavy tail modelling. An important contribution of this kind in non-life insurance rate-making is \cite{frees2008hierarchical}, in which   the four-parameter GB2 family was used as a response distribution. Here we can also refer to \cite{gunduz2016exponentiated} for an exponentiated Fr\' echet regression model,   \cite{beirlant1998burr} for Burr regression,
 \cite{frangos2004modelling} for an exponential-inverse Gaussian regression, \cite{gomez2013gamma}
for a gamma-generalized inverse Gaussian regression,
and \cite{stasinopoulos2007generalized} for a log-normal regression model.
In this paper we propose  parametric regression modelling using an appropriate subfamily of the GB2 family which appears still to be able to fit tail and body of heavy tailed loss data appropriately.

The starting point for heavy tailed regression modelling is to find an appropriate heavy-tailed (univariate) distribution. Extreme Value Theory (EVT) as discussed for instance in \cite{embrechts2013modelling} and \cite{albrecher2017}, provides procedures to fit  the simple Pareto or the generalized Pareto (GP) distribution to data in excess of an appropriate high threshold.  In so-called splicing or composite
models, in order to extend such tail fits to the modal part of the data, one can combine a tail fit and a distribution modelling the loss data below the threshold,  see e.g. \cite{cooray2005modeling}, \cite{scollnik2007composite}, \cite{scollnik2012modeling}, \cite{bakar2015modeling} and \cite{del2017full}. In \cite{grun2019extending}
a comprehensive analysis is provided for the Danish fire losses data set by evaluating
256 composite models derived from 16 parametric distributions that are
commonly used in actuarial science.
In \cite{reynkens2017modelling} the modal part fit is established using a mixed Erlang distribution, which can also be adapted to censored data. This then reduces the problem of selecting a specific parametric modal part component.

Another approach consists of using transformation methods, such as the
log-skew-normal distribution \citep{azzalini2002log},
the log-skew-t distribution \citep{landsman2016tail}, the generalized logMoyal distribution \citep{bhati2018generalized} and transformed beta (generalized beta of
the second kind, GB2) distributions \citep{cummins2004risky}, among others.
In particular, the four-parameter GB2 family is a very useful tool in the actuarial literature in the studies of the insurance loss distribution \citep{frees2008hierarchical,shi2015private} and includes many of the
aforementioned distributions as special or limiting cases. In \cite{tencaliec2019} a transformation model, termed the  extended generalized Pareto distribution, is presented which is in compliance with extreme value theory for both small and large values.

Finite/infinite mixture models constitute another method which deals with modelling heavy-tailed losses. Mixture models can also be used to capture the
heterogeneity in the data and allow for the mixture components
to represent groups in the population.
This approach has appeared in several publications in non-life actuarial literature. For example,
\cite{bernardi2012skew} proposes finite mixture of Skew Normal distributions in the framework of Bayesian analysis.
\cite{verbelen2015fitting} develops finite mixtures of Erlang distributions and adopt the EM algorithm to estimate the model.
\cite{gomez2013gamma} proposes a gamma mixture with the generalized inverse Gaussian distribution to fit the well known Danish fire data set.
 \cite{miljkovic2016modeling} extended the distribution of finite mixture models to more general forms, such as the Burr, Gamma, Inverse Burr, Inverse Gaussian, lognormal, Weibull, and GB2 distribution \citep{chan2018modelling}. In addition,
 \cite{li2016bayesian} proposes to use the mixture model to estimate the catastrophic model, and apply Bayesian method to calculate the Value-at-Risk and Expected Shortfall probability.
 Recently, \cite{punzo2018compound} proposes a three-parameter
compound (mixed) distribution in order to take care of specifics such as
uni-modality, hump-shaped, right-skewed and heavy tails. However, the resultant density obtained by \cite{punzo2018compound} may not always have closed form expressions which
make the estimation more cumbersome.

The use of heavy tailed transformation models, splicing models or finite mixture models in a regression setting have not been fully established yet. Extreme value regression models did take off when confined to tail parameter estimation only, starting with the seminal work in \cite{davison1990}.
 Motivated by the recent publications \cite{punzo2018compound} and \cite{bhati2018generalized},
we propose a gamma mixture of the recent parametric log-Moyal distribution proposed by \cite{bhati2018generalized}. We hence add one extra  parameter to the log-Moyal model that allows to model the extreme heavy-tailed data and is flexible in regression modelling. The original Moyal distribution was proposed in a 1955 paper by physicist J.E. Moyal in quantum mechanics describing the energy lost by a fast charged particle during ionization.

 We further show some important features such as closed form expressions for the probability density function, moments, risk measures, $\ldots$
This new class constitutes a special case of the four parameter GB2 distribution setting  one of the related shape parameters to 0.5.
As illustrated below, this new generalized log-Moyal gamma (GLMGA) distribution can be usefully applied in loss modelling.
The advantages of the proposed GLMGA model include the following:
(1) the class has a power law tail, suitable for modelling heavy tailed data;
(2) the model can provide a suitable fit for the
entire range of data apart from the tails.
To the best of our knowledge, Moyal related distributions or any of its
extensions have not been explored except the generalized log-Moyal distribution $\text{GlogM}(\theta, \sigma)$ and the beta-Moyal distribution
proposed in \cite{bhati2018generalized} and \cite{cordeiro2012beta}.

The remainder of the article is structured as follows.
In Section \ref{section: proposed model} we provide a brief
summary of the GlogM distribution, introduce the GLMGA distribution and study some properties, such as its tail behaviour and risk measure expressions.
Regression modelling is discussed in Section  \ref{section:regression} and studied in Section \ref{section:simulations} through simulations.
To illustrate its practical use, in Section \ref{section:empirical}, we fit the GLMGA to a fire claim data set and apply
the GLMGA regression procedure to an earthquake loss data set from China, comparing with fits based on models from literature.
Finally, some conclusions, along
with future possible extensions, are drawn in Section \ref{section:conclusion}.
The code that was used to analyse the data can be obtained from
\url{https://github.com/lizhengxiao/GLMGA-model} .

\section{The generalized log-Moyal gamma distribution}\label{section: proposed model}

The generalized log-Moyal distribution, to be denoted by $\text{GlogM}(\theta, \sigma)$, was introduced by \cite{bhati2018generalized}.
$\text{GlogM}(\theta, \sigma)$ is generalization of the continuous Moyal distribution \citep{moyal1955} using the transformation method,  exhibiting unimodality and  right skewness with the right tail being heavier than the exponential model.
The density function and distribution function are given  by
\begin{align}
   {f}(y)&=\frac{\sqrt{\theta}}{\sqrt{2\pi }\sigma }\ {{\left( \frac{1}{y} \right)}^{\frac{1}{2\sigma }+1}}{{\exp}{\left[-\frac{\theta}{2}{{\left( \frac{1}{y} \right)}^{1/\sigma}}\right]}},\label{GlogM:density}\\
   F(y)&=\text{erfc}\left[\left(\frac{\theta}{2y^{1/\sigma}}\right)^{\frac{1}{2}}
   \right], \; y>0, \quad \sigma >0 \quad \theta>0,
   \label{GlogM:distribution}
\end{align}
where $\text{erfc}(\cdot)$ is the complementary error function given by $\text{erfc}(z)=\frac{2}{\sqrt{\pi}}\int_z^{\infty}\exp(-t^2)dt$.
\\
The $r^{th}$ moment  of $Y\sim\text{GlogM}(\theta,\sigma)$
is given by
\[
\mathbb{E}(Y^r)=\left(\frac{\theta}{2}\right)^{r\sigma}\frac{\Gamma(\frac{1}{2}-r\sigma)}{\Gamma(\frac{1}{2})}, \quad \quad \sigma <\frac{1}{2r},
\]
so that the mean and variance of the $\text{GlogM}(\theta,\sigma)$ distribution exist if and
only $\sigma<\frac{1}{2}$ and $\sigma<\frac{1}{4}$ respectively:
\begin{align*}
\mathbb{E}(Y)&=\left(\frac{\theta}{2}\right)^{\sigma}\frac{\Gamma(\frac{1}{2}-\sigma)}{\Gamma(\frac{1}{2})},\quad\quad \sigma<\frac{1}{2},\\
\text{Var}(Y)&=\left(\frac{\theta}{2}\right)^{2\sigma}\frac{\Gamma(\frac{1}{2})\Gamma(\frac{1}{2}-2\sigma)-\Gamma^2(\frac{1}{2}-\sigma)}{\Gamma^2(\frac{1}{2})}
,\quad\quad \sigma<\frac{1}{4}.
\end{align*}

\vspace{0.3cm}
We here apply the infinite mixture approach improving on the modelling of heavy-tailed  data, using the classical gamma distribution as the mixture distribution.
We further derive closed form expressions of some important features such as   cumulative distribution function, moments, risk measures, etc. of the resulting three-parameter distribution.
\begin{definition} A positive random variable $Y$ follows a generalized log-Moyal gamma distribution if it admits the stochastic representation
\[
Y|\Theta \sim \text{GlogM}(\Theta,\sigma)\quad \text{and}\quad \Theta \sim \text{Gamma}(a, b),
\]
where $\text{Gamma}(a, b)$ refers to a gamma distribution with density
\[
p(\theta) =\frac{{b }^{a }}{\Gamma \left( a\right)}\theta^{a-1}\exp{\left(-b \theta\right)}
\]
with $a,b>0$.
\label{def GLMGA}
\end{definition}

\vspace{0.3cm}
Definition \ref{def GLMGA} leads to the $\text{GLMGA}(\sigma,a,b)$ density by integrating out $\Theta$:
\begin{align}
f\left(y \right)&=\frac{b^a}{\sqrt{2}\sigma B\left(a,\frac{1}{2}\right)}\frac{y^{-\left(\frac{1}{2\sigma}+1\right)}}{\left(\frac{1}{2}y^{-\frac{1}{\sigma}}+b\right)^{a + \frac{1}{2}}}
\label{pdf:LMGA}\\
&=\frac{(2b)^a}{\sigma B(a,\frac{1}{2})}\frac{y^{-\left(\frac{1}{2\sigma}+1\right)}}{\left(y^{-\frac{1}{\sigma}}+2b\right)^{a + \frac{1}{2}}},
\end{align}
for $y>0$, with $\sigma>0, a>0, b>0$. Here $B(m,n)=\int_0^{1}t^{m-1}(1-t)^{n-1}dt$ is the beta function.\\ 

We next list some expressions and properties of the $\text{GLMGA}(\sigma,a,b)$ distribution. Proofs are deferred to the Appendix.
\begin{proposition}
The cumulative distribution function (cdf) $F$ and quantile function $F^{-1}$ of
the generalized log-Moyal gamma distribution are given by
\begin{align}
F\left(y\right)&=1-I_{\frac{1}{2},a}\left(\frac{y^{-1/\sigma}}{y^{-1/\sigma}+2b}\right),
\label{cdf:GLMA}\\
F^{-1}(u)&=(2b)^{-\sigma}\left[\frac{I^{-1}_{\frac{1}{2},a}(1-u)}{1-I^{-1}_{\frac{1}{2},a}(1-u)}\right]^{-\sigma},
\label{qf:GLMA}
\end{align}
where $I_{m,n}(t)$ is the beta cumulative distribution function with two positive parameters $m$ and $n$, and $I_{m,n}^{-1}(t)$ is its inverse.
\end{proposition}

Concerning the relationship of $\text{GLMGA}(\sigma,a,b)$  with some
 other well known families of distributions, we can state the following.
\begin{itemize}
\item The density of the generalized beta distribution of the second kind (GB2) is given by
\begin{equation}
f\left(y \right)=\frac{\abs{p}}{B(\nu,\tau)y}\frac{\mu^{p \tau}y^{p\nu}}{(y^p+\mu^p)^{\nu+\tau}},
\label{pdf:GB2}
\end{equation}
where $\tau,\nu >0, \mu>0$ and $p\in \mathbb{R}$.
The substitution  $(\tau=a,\mu=(2b)^{-\sigma},\nu=\frac{1}{2},p=-\frac{1}{\sigma})$ yields the $\text{GLMGA}(\sigma,a,b)$ distribution. More applications of the univariate GB2 and its extension can be found  in \cite{mcdonald1990regression}, \cite{mcdonald1991option}, \cite{yang2011generalized} , \cite{jeong2019bayesian} and \cite{dong2013bayesian}.
\item If  $X$ follows the standard Moyal distribution \citep{moyal1955, bhati2018generalized} with the density function $g(x)=\frac{1}{\sqrt{2\pi}}\exp\left[-(x+\exp(-x))/2\right]$ and  $Y$ is gamma$(a,1)$ distributed  with shape parameter $a$ and unit scale parameter, then $Z=(Y/b)^\sigma\exp(\sigma X)\sim \text{GLMGA}(\sigma,a,b)$.
\item If  $X$ follows the half-normal $(0,\sigma^2)$ distribution \citep{bhati2018generalized} and  $Y$ is gamma$(a,1)$ distributed, then  $Z=\left(Y/{(bX^2)}\right)^\sigma\sim \text{GLMGA}(\sigma,a,b)$.
\item If  $X$ and $Y$ are independent gamma distributed with common unit scale parameter  and shape parameters $\frac{1}{2}$ and $a$ \citep{cummins1990applications}, then  $Z=\left(Y/(2bX)\right)^\sigma \sim\text{GLMGA}(\sigma, a, b)$.
\item If $X\sim\text{GLMGA}(\sigma,a,b)$, then  $Y=kX\sim\text{GLMGA}(\sigma,a,bk^{-1/\sigma})$.
\item  for $\sigma=\frac{1}{2}$ and $a=b$, the GLMGA density (\ref{pdf:LMGA}) reduces to the inverse folded-t distribution \citep{brazauskas2011folded} with unit scale parameter  and degrees of freedom $2a=1,2,3,\dots$.
    Moreover, for $a=b=\frac{1}{2}$, the inverse folded-t distribution reduces to the inverse standard folded Cauchy distribution.
\item for $a\to\infty$ the density $f_{\sigma,a,b}$ of the $\text{GLMGA}(\sigma, a, b)$ distribution given in  (\ref{pdf:LMGA}) of  generalized inverse gamma distribution
with shape parameters $\frac{1}{2}$ and  $\frac{1}{\sigma}$, and scale parameter $\phi = (2b/a)^{\sigma}$
\citep{stacy1962generalization,mead2015generalized}:
\begin{equation}
\lim_{a \to \infty} f_{\sigma,a,b}(y) =
\frac{1}{\sigma\sqrt{\pi}\phi^{\frac{1}{2\sigma}}}y^{-\frac{1}{2\sigma}-1}\exp\left[-\left(\phi y\right)^{-\frac{1}{\sigma}}\right].
\label{lim_density}
\end{equation}
 Moreover, for $\sigma=\frac{1}{2}$ and $a\to\infty$ the $\text{GLMGA}(\sigma, a, b)$ density in  (\ref{pdf:LMGA}) reduces to the  inverse half-normal distribution.
\end{itemize}


\vspace{0.3cm}
{\bf The $r^{th}$ moments}  of the GLMGA distribution are defined for $r\sigma<\frac{1}{2}$:
\begin{equation}
 \mathbb{E}(Y^r)=\frac{(2b)^{-\sigma r}B(\frac{1}{2}-r\sigma,a+r\sigma)}{B(\frac{1}{2}, a)}.
 \label{moments:GLMGA}
\end{equation}
In particular, the mean and variance are given by
\begin{align*}
 \mathbb{E}(Y)&=\frac{(2b)^{-\sigma}B(\frac{1}{2}-\sigma,a+\sigma)}{B(\frac{1}{2}, a)}, \quad \sigma <\frac{1}{2},\\
\text{Var}(Y)&=\frac{(2b)^{-2\sigma}}{B(\frac{1}{2},a)^2}\left[
B(\frac{1}{2}-2\sigma,a+2\sigma)B(\frac{1}{2},a)-B(\frac{1}{2}-\sigma,a+\sigma)^2
\right], \quad \sigma <\frac{1}{4}.
\end{align*}
 Concerning the  $r^{th}$ incomplete (conditional) moments of  the GLMGA distribution, given $y\le u$ and $y>u$, one finds
\begin{align}
\mathbb{E}(Y^r|Y\le u)&=\frac{1}{F_Y(u)}\int_{0}^{u}y^r f(y)dy \nonumber\\
&=\mathbb{E}(Y^r)
\frac{1-I_{\frac{1}{2}-r\sigma,a+r\sigma}\left[\frac{u^{-1/\sigma}}{u^{-1/\sigma}+2b}\right]}
{1-I_{\frac{1}{2},a}\left[\frac{u^{-1/\sigma}}{u^{-1/\sigma}+2b}\right]},
\label{incomplete_moment_bottom}
\end{align}
and
\begin{align}
\mathbb{E}(Y^r|Y>u)&=\frac{1}{1-F_Y(u)}\int_{u}^{+\infty}y^r f(y)dy \nonumber\\
&=\mathbb{E}(Y^r)
\frac{I_{\frac{1}{2}-r\sigma,a+r\sigma}\left[\frac{u^{-1/\sigma}}{u^{-1/\sigma}+2b}\right]}
{I_{\frac{1}{2},a}\left[\frac{u^{-1/\sigma}}{u^{-1/\sigma}+2b}\right]}.
\label{incomplete_moment_top}
\end{align}


\vspace{0.4cm}
The  $\text{GLMGA}(\sigma, a, b)$ is {\bf unimodal}.
Equating the derivative of the logarithm of the density (\ref{pdf:LMGA}) to zero, one obtains the mode $y_0$ of the $\text{GLMGA}(\sigma, a, b)$ distribution
\begin{equation}
y_0 = \left(\frac{b+2b\sigma}{a-\sigma}\right)^{-\sigma}.
\end{equation}
Figure \ref{fig:mean,median and mode} represents the skewness of the $ \text{GLMGA}(\sigma, a, b)$ model, from which the order mean $>$ median $>$ mode can be observed.
\begin{figure}[htbp]
	\centering
	\includegraphics[scale=0.5]{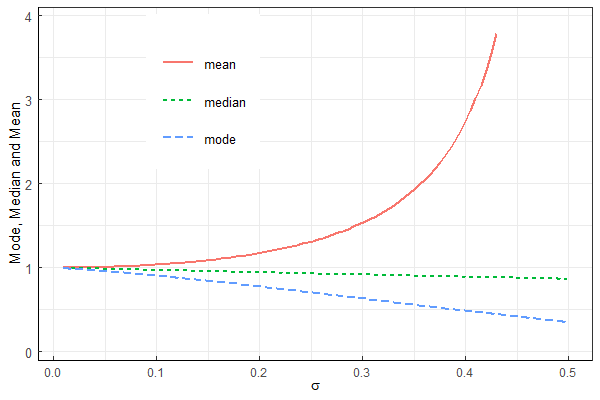}
	\caption{Pattern of mean, median and mode of the generalized log-Moyal gamma distribution for $a = 1, b = 2$ and different values of $\sigma$.}
	\label{fig:mean,median and mode}
\end{figure}


\begin{figure}[htbp]
	\centering
	\includegraphics[width=0.5\textwidth]{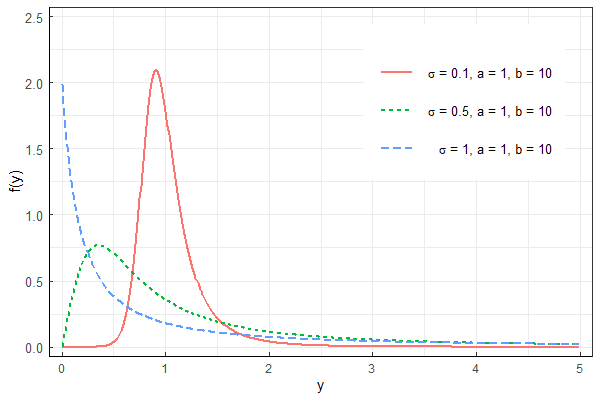} \\
	\includegraphics[width=0.5\textwidth]{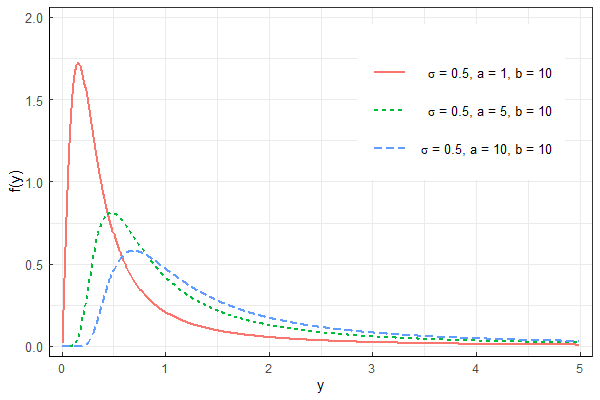}\\
	\includegraphics[width=0.5\textwidth]{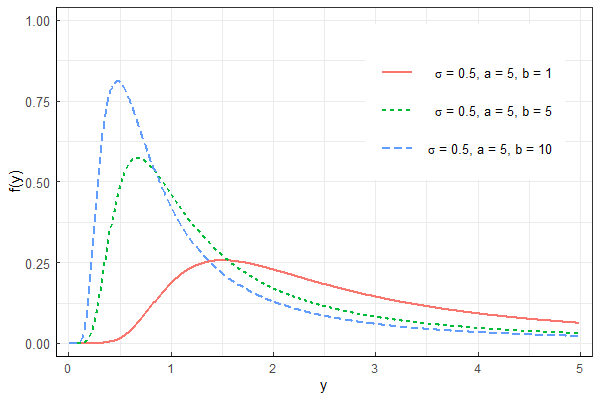}
	\caption{Probability density function across parameter $\sigma, a, b$ in the generalized log-Moyal gamma distribution.}
	\label{fig:pdf}
\end{figure}

Figure \ref{fig:pdf} demonstrates how the density
function of the generalized log-Moyal gamma distribution changes when one or two of the three parameters vary while the others are fixed. It can be observed, in all cases, that the model has positive skewness.

\vspace{0.3cm}
On the other hand,
 the higher $\sigma$, the heavier the right tail of the distribution. In fact, the GLMGA distribution is of Pareto-type, i.e. the tail function $\bar{F}=1-F$ is regularly varying at infinity:
\[
\bar{F}(y)=y^{-1/\xi} L(y), \quad \quad y\uparrow \infty,
\]
where $ L(y)$ is slowly varying at infinity
\[
\lim_{t \to \infty}\frac{L(ty)}{L(t)}=1, \mbox{ for every } y >0.
\]
Then $\xi$ is called the extreme value index, or equivalently $1/\xi$ is termed the Pareto index.
See for instance \cite{embrechts2013modelling} or \cite{beirlant2004} for more details.
Indeed, for the GLMGA distribution we find that
\begin{equation}
\bar{F}(y)= Cy^{-1/(2\sigma)}\left\{1 + Dy^{-1/\sigma}\left(1+o(1)\right)\right\}, \quad \quad y\to \infty,
\label{Patype}
\end{equation}
with $C=\frac{2}{(2b)^{1/2}B(a,\frac{1}{2})}>0$ and $D=-\frac{2a+1}{12b}$.
Note that the extreme value index equals $2\sigma$ in this case, and that
$
L(y)=C\left\{1+Dy^{-\frac{1}{\sigma}}\left(1+o(1)\right)\right\} \;\; (y \to \infty).
$

\vspace{0.3cm}
We end this section computing some important {\bf risk measure} expressions for the GLMGA distribution.
The Value-at-Risk (VaR) was already given in \eqref{qf:GLMA}:
\begin{equation}
\text{VaR}_p(Y)= F^{-1}(p) =(2b)^{-\sigma}\left[\frac{I^{-1}_{\frac{1}{2},a}(1-p)}{1-I^{-1}_{\frac{1}{2},a}(1-p)}\right]^{-\sigma}.
\end{equation}

A closed form expression can also be obtained for the Tail-Value-at-Risk, denoted as
$\text{TVaR}_p(Y)$:
\begin{align}
\text{TVaR}_p(Y)&=\mathbb{E}(Y|Y>\pi_p)=\frac{\int_{\pi_p}^{\infty}yf(y)dy}{1-F(\pi_p)}\nonumber\\
&=\mathbb{E}(Y)
\frac{I_{\frac{1}{2}-\sigma,a+\sigma}\left[I_{\frac{1}{2},a}^{-1}(1-p)\right]}
{1-p}\nonumber\\
&=\frac{(2b)^{-\sigma}B(\frac{1}{2}-\sigma,a+\sigma)}{B(\frac{1}{2}, a)}
\frac{I_{\frac{1}{2}-\sigma,a+\sigma}\left[I_{\frac{1}{2},a}^{-1}(1-p)\right]}
{1-p}, \mbox{ if } \sigma <1/2.
\end{align}
The net premium of  an (unbounded) Excess-of-Loss reinsurance contract where the reinsurer pays the amount in excess of a line $R$, is given by
\begin{align}
\mathbb{E}\left[(Y-R)_{+}\right]&=\mathbb{E}\left[\max(0,Y-R)\right]\nonumber\\
&=\int_{R}^{\infty}yf_{Y}(y)dy-R\left[1-F_{Y}(R)\right]\nonumber\\
&=\left(\mathbb{E}(Y)-R\right)I_{\frac{1}{2},a}\left[\frac{R^{-1/\sigma}}{R^{-1/\sigma}+2b}\right].
\label{excess:GLMGA}
\end{align}
Finally, the mean excess function of the GLMGA distribution is given by
\begin{equation}
\mathbb{E}(Y-u|Y>u)
=\left[\mathbb{E}(Y)-u\right]
\frac{I_{\frac{1}{2}-\sigma,a+\sigma}\left[\frac{u^{-1/\sigma}}{u^{-1/\sigma}+2b}\right]}
{I_{\frac{1}{2},a}\left[\frac{u^{-1/\sigma}}{u^{-1/\sigma}+2b}\right]}.
\label{MEF}
\end{equation}
In view of \eqref{Patype} and (3.4.12) in  \cite{albrecher2017} we then have that
\[
\lim_{u \to \infty} {\mathbb{E}(Y-u|Y>u) \over u} =
{2\sigma \over 1-2\sigma}, \mbox{ if } \sigma <1/2.
\]

\section{A regression GLMGA model}\label{section:regression}
Whereas in classical regression analysis the dependence of the response variable
on the covariate(s) is modelled via the conditional
mean of the response variable, in situations with
 a heavy tailed response variable
 (some of) the response model parameters  are treated directly as functions of the
covariate, among others because the mean of the distribution may not always exist. \cite{dong2013bayesian}
discussed this regression approach assuming
that the response variable follows a GB2 distribution which nests various
distributions with light and heavy tails, to facilitate accurate loss reserving in insurance applications, and recently \cite{bhati2018generalized} considered a GlogM regression model. Here we assume that the response variable follows the GLMGA distribution  and propose the parameters $\sigma$ and/or $b$ to be modelled as a function of the explanatory variables.
In order to avoid boundary problems in optimization, we consider a log link function obtaining real-valued  parameters:

\begin{align}
Y_{i}|\bm{x}_{\sigma,i},\bm{x}_{b,i}&\sim \text{GLMGA}(\sigma_i,a,b_i), \nonumber\\
&\log(\sigma_{i})=\bm{x}^{T}_{\sigma,i}\bm{\beta},\\
&\log(b_{i})=\bm{x}^{T}_{b,i}\bm{\alpha},\nonumber\\
&\log(a)=\eta,\nonumber
\end{align}
where $\bm{x}_{\sigma,i}^{T}=(1,x_{\sigma,i1},\dots,x_{\sigma,ik})$, $\bm{x}_{b,i}^{T}=(1,x_{b,i1},\dots,x_{b,ik})$ denote the vector of covariates and $\bm{\beta}=(\beta_0,\beta_{1},\dots,\beta_{k})^{T}$, $\bm{\alpha}=(\alpha_0,\alpha_{1},\dots,\alpha_{k})^{T}$  the vector of coefficients.

We consider estimation of the model parameters using maximum likelihood estimation (MLE). Denoting the $n$ independent observations from the GLMGA distribution with unknown parameter vector $(\bm{\beta},\bm{\alpha},\eta)$ by  $\bm{y}=(y_1,\dots,y_n)^{T}$, the  log-likelihood
function equals
\begin{align}
\ell(\bm{\beta},\bm{\alpha},a)=&a\sum_{i=1}^{n}\log(b_i)-\frac{n}{2}\log(2)
-n\log B(a,\frac{1}{2})\nonumber\\
&-\sum_{i=1}^{n}\frac{\log(y_i)}{2\sigma_i+1}-(a+\frac{1}{2})\sum_{i=1}^{n}\left[\frac{1}{2}y^{-\frac{1}{\sigma_i}}+b_i
\right].
\label{loglike:LMGA}
\end{align}
We use the optim() function in R which uses the Nelder-Mead opimization method. The asymptotic variance-covariance matrix of the estimators is computed
as  the inverse of the observed Fisher information matrix.

For assessment of the regression model we use  randomized quantile residuals  defined by
$r_{i}=\Phi^{-1}\left[F(y_{i};\hat{\bm{\beta}},\hat{\bm{\alpha}},\hat{\eta})\right]$ ($i=1,\ldots,n$),
where $\Phi\left(\cdot\right)$ is the cdf of the standard normal distribution and $F$ denotes the cdf of the GLMGA model as given in \eqref{cdf:GLMA}, and parameter $\hat{a}=\exp(\hat{\eta})$. The distribution of $r_{i}$ converges to standard normal if $(\hat{\bm{\beta}},\hat{\bm{\alpha}},\hat{\eta})$ are consistently estimated, see \cite{dunn1996randomized}, and hence a normal QQ-plot of the $r_i$ should follow the 45 degree line for the regression application to be relevant.

\section{Simulation study}\label{section:simulations}
\begin{figure}[htbp]
	\centering
	\includegraphics[scale=0.75]{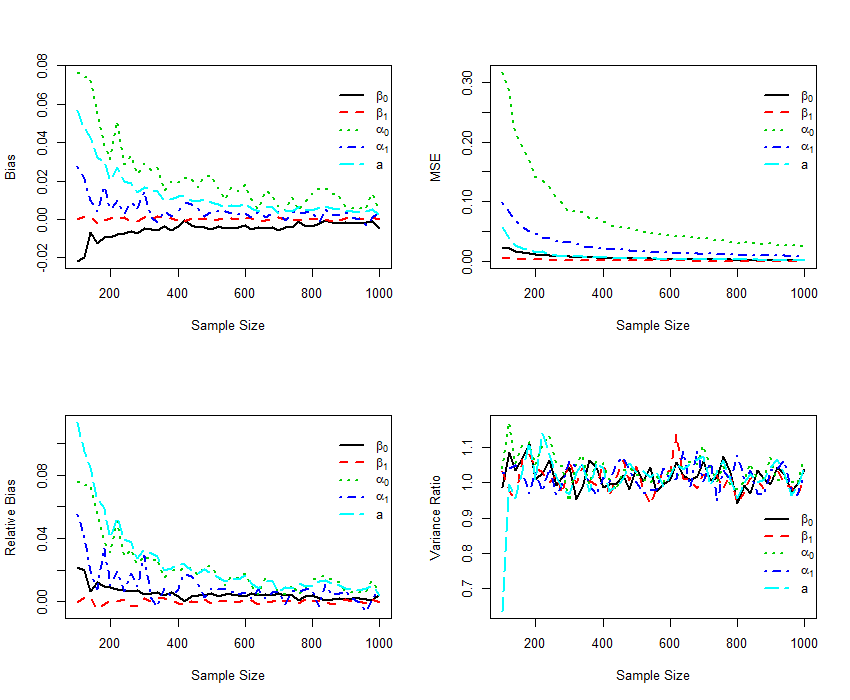}
	\caption{Bias (top left), and mean squared error (top right), relative bias error (bottom left) and  the ratio of the sample variance to the asymptotic
variance (bottom right) of the parameter estimates $\beta_0$ (black and solid), $\beta_1$ (red and dashed), $\alpha_0$ (green and dotted), $\alpha_1$ (darkblue and dotdash) and $a$ (cyan and longdash) in case of $(\beta_0,\beta_1,\alpha_0,\alpha_1,a) = (-1,0.5,1,0.5,0.5)$. The sample size is from $n = 100$ to $n =1000$. The plots are obtained by averaging
out over 2000 samples.}
	\label{fig: bias+mse}
\end{figure}

\begin{figure}[htbp]
	\centering
	\includegraphics[scale=0.7]{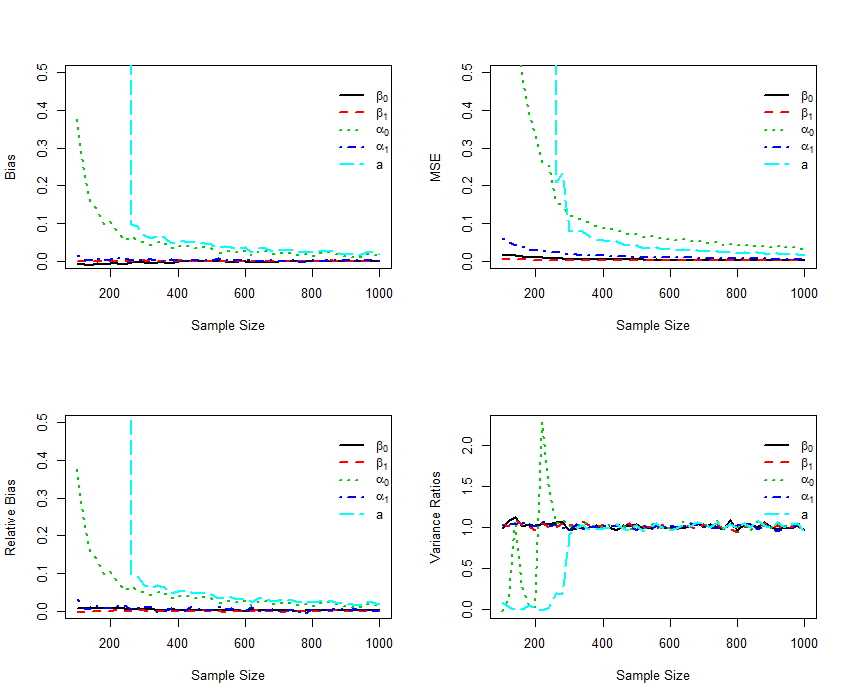}
	\caption{Bias (top left), and mean squared error (top right), relative bias error (bottom left) and  the ratio of the sample variance to the asymptotic
variance (bottom right) of the parameter estimates $\beta_0$ (black and solid), $\beta_1$ (red and dashed), $\alpha_0$ (green and dotted), $\alpha_1$ (darkblue and dotdash) and $a$ (cyan and longdash) in case of $(\beta_0,\beta_1,\alpha_0,\alpha_1,a) = (-1,0.5,1,0.5,1)$. The sample size is from $n = 100$ to $n =1000$. The plots are obtained by averaging
out over 2000 samples.}
	\label{fig: bias+mse2}
\end{figure}

In this section, we first perform a simulation study to check the accuracy of the ML estimators  based on the proposed GLMGA regression model.
We generated 2000 data sets of sizes from $n =200$ to  $n=1000$ from GLMGA regression model with $k=1$, $\bm{\beta}=(-1,0.5)$,
$\bm{\alpha}=(1,0.5)$,  $\bm{x}_{\sigma,i}=(1,x_{i,1} )$, $\bm{x}_{b,i}=(1, x_{i,2})$
for $a=0.5,1$, and with the covariates $x_{i,1}$ and $x_{i,2}$ being generated from the standard normal distribution.
Figure \ref{fig: bias+mse}-\ref{fig: bias+mse2} report the absolute bias, relative bias error (that is, sample mean over the true value -1), the ratio of the sample variance to the asymptotic variance, and the mean squared error (MSE).

\begin{figure}[htbp]
	\centering
	\includegraphics[scale=0.55]{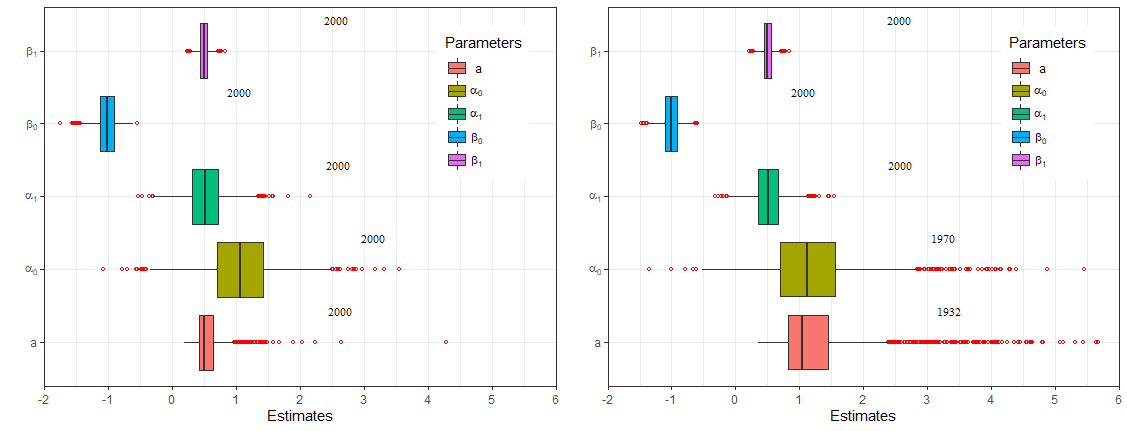}
	\caption{
Boxplots of the parameter estimates from 2000 GLMGA simulated samples of size $n=100$.
{\it Left}:  results for  $(\beta_0,\beta_1,\alpha_0,\alpha_1,a) = (-1,0.5,1,0.5,0.5)$.
{\it Right}: results for   $(\beta_0,\beta_1,\alpha_0,\alpha_1,a) = (-1,0.5,1,0.5,1)$.
The number of included samples are given next to each boxplot. }
	\label{fig: boxplot}
\end{figure}

Figures \ref{fig: bias+mse} and \ref{fig: bias+mse2}
show that model parameters $\alpha_0, \alpha_1, a$ are slightly overestimated.
The parameter $a$ is unstable
when the sample size is small.
As the sample size increases, the estimators close up to the true values,
with smaller bias, variance ratios, relative bias errors and MSE.
Whereas  $(\beta_0,\beta_1)$ are estimated accurately even with small sample sizes, 
the estimators of $\alpha_0,\alpha_1$ show more instability when the sample size is small. The bias, variance and MSE have increased with the  larger value of $a$.

In order to study the behaviour of the estimation technique for small sample sizes, in Figure \ref{fig: boxplot} we present the boxplots of the parameter estimates from 2000 Monte Carlo simulations with small sample size $n = 100$.
In the case  $a=1$, the median estimate of  $a$ is 1.1 while the mean is 4810 with maximum 1707622, and 3.40\% of the  2000 sample estimates  is situated  above the upper limit of this figure.
Moreover, the median  $\alpha_0$ estimate is 1.12, while the mean is 1.37 with maximum equal to 16.15, and only 1.65\% of the 2000 sample estimates lie  above the upper figure limit.

To demonstrate the approximate normality of the estimators of GLMGA regression parameters,
 in Figures \ref{fig:size200-a=0.5}-\ref{fig:size1000-a=1} in Appendix C, the normal QQ-plots of the estimated parameters are presented
for sample sizes $n=200, 2000$ and $a=0.5, 1$.
The dashed lines indicate
95\% confidence intervals\footnote{Confidence intervals are calculated using $+/-k$\%, where
$ k = 89.5 / (\sqrt{n} \times (1- 0.01 /\sqrt{n} + 0.85/n))$. Gives a 95\% asymptotic band based on the Kolmogorov-Smirnov statistic, see \cite{doksum1976plotting}.}.
The match between the
theoretical and empirical quantiles suggest acceptable estimation results, while some lack of  normality for estimation of $a$ can be observed
when the true value $a=1$  and the sample size is small.

\newpage
\section{Real-data illustrations}\label{section:empirical}
In this section we will illustrate the proposed method with the two practical examples
introduced in Section \ref{section: proposed model}.
\subsection{
Application of the univariate GLMGA distribution to the fire claim data set}
As a first example, we fit the univariate GLMGA distribution to the fire claim data at a major university reported in \cite{cummins1990applications}, which consists of 80 fire claims.
The data cover several years and have been adjusted to a
common time point using a claims cost index maintained by
the university from which the claims were obtained.
The data are described in more detail in \cite{cummins1978comparative}.

\begin{figure}[htbp]
	\centering
	\includegraphics[scale=0.6]{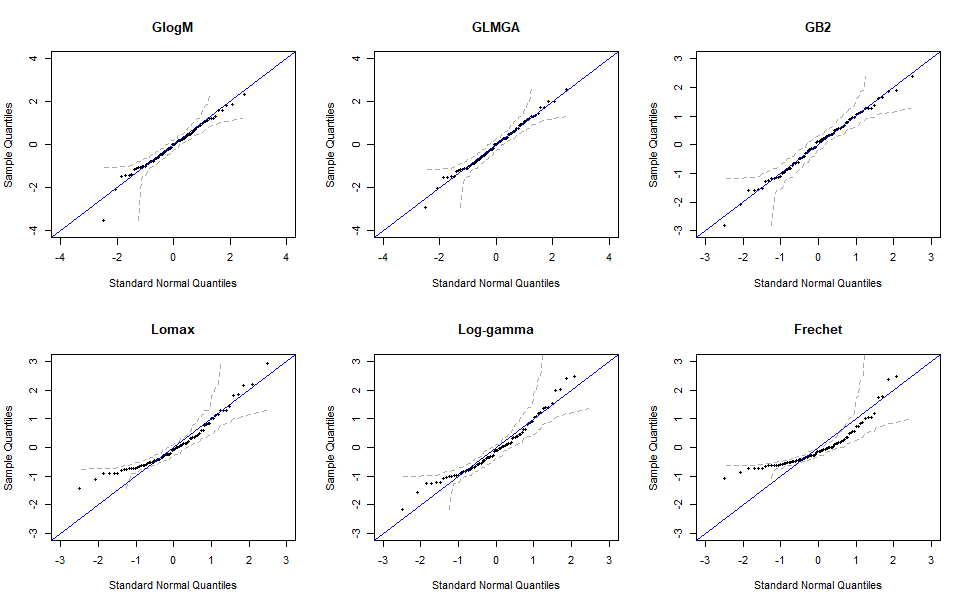} 
	\caption{Normal QQ-plots of the quantile residuals for the fire claim data set based on six heavy-tailed distributions.}
	\label{fig:QQ-plot-data I}
\end{figure}

We compare the   GLMGA model with five other competitive heavy tailed distributions, namely, log gamma, Fr\'echet, GB2, log-Moyal, and Lomax.
In Table \ref{tab:estimates - data I} we provide the estimates, log-likelihood values
(LL), as well as the Akaike Information Criterion (AIC) and the Bayesian Information Criterion (BIC) values, defined as $\text{AIC}=-2\ell+2p$ and $\text{BIC}=-2\ell+p\log n$, where $\ell$ is the log-likelihood value, $p$ is the number of model parameters, and
$n$ is the number
of observations.
We further compute the bootstrap p-values for some goodness-of-fit tests, namely, Kolmogorov-Smirnov (KS), Cramer-von Mises
(CvM) and Anderson-Darling (AD) tests.
It is clear from Table \ref{tab:estimates - data I} that the GlogM, GLMGA and GB2 provide a better fit as they have the highest log-likelihood value and the minimum AIC and BIC
values. Also, the bootstrap p-values for KS, CvM and AD tests
for the GlogM, GLMGA and GB2 distribution rank highest.
Further, the normal QQ-plots based on the quantile residuals given in Figure \ref{fig:QQ-plot-data I}
also indicate that  the GlogM, GLMGA and GB2 provide a better fit. To some extent the degree of linearity is captured using the correlation coefficient of the normal QQ-plots $R$ in Table \ref{tab:estimates - data I}.

\begin{table}[htbp]
  \centering
  \caption{Data analysis for the fire claim data set}
  \begin{tabular*}{\hsize}{@{}@{\extracolsep{\fill}}ccllccccccc@{}}
    \hline\hline
    \multicolumn{1}{l}{Distribution} & \multicolumn{1}{l}{ \#Par.} & \multicolumn{2}{c}{Estimates} & \multicolumn{1}{l}{LL} & \multicolumn{1}{l}{AIC} & \multicolumn{1}{l}{BIC} & \multicolumn{1}{l}{KS} & \multicolumn{1}{l}{AD} & \multicolumn{1}{l}{CvM}
    & \multicolumn{1}{l}{R}  \\
    \hline
    \multirow{2}[0]{*}{GlogM} & \multirow{2}[0]{*}{2} & $\hat{\sigma}$ & 0.667 & \multirow{2}[0]{*}{-786.22} & \multirow{2}[0]{*}{1576.4} & \multirow{2}[0]{*}{\bf 1581.2} & \multirow{2}[0]{*}{0.988} & \multirow{2}[0]{*}{0.973} & \multirow{2}[0]{*}{0.992} & \multirow{2}[0]{*}{0.989} \\
          &       & $\hat{\mu}$    & 1662.371 &       &       &       &       &       & & \\
          \hline
    \multirow{4}[0]{*}{GB2} & \multirow{4}[0]{*}{4} & $\hat{\mu}$    & 1097.730 & \multirow{4}[0]{*}{-784.61} & \multirow{4}[0]{*}{1577.2} & \multirow{4}[0]{*}{1586.8} & \multirow{4}[0]{*}{0.958} & \multirow{4}[0]{*}{\bf 0.998} & \multirow{4}[0]{*}{0.988} &  \multirow{4}[0]{*}{\bf 0.998} \\
          &       & $\hat{\sigma}$ & 4.043 &       &       &       &       &       &  &\\
          &       & $\hat{\nu}$    & 0.832 &       &       &       &       &       & & \\
          &       & $\hat{\tau}$   & 0.182 &       &       &       &       &       & & \\
          \hline
    \multirow{3}[0]{*}{GLMGA} & \multirow{3}[0]{*}{3} & $\hat{\sigma}$ & 0.580 & \multirow{3}[0]{*}{\bf -784.97} & \multirow{3}[0]{*}{\bf 1575.9} & \multirow{3}[0]{*}{1583.1} & \multirow{3}[0]{*}{\bf 0.999} & \multirow{3}[0]{*}{0.996} & \multirow{3}[0]{*}{\bf 0.995} & \multirow{3}[0]{*}{ 0.996} \\
          &       & $\hat{a}$    & 0.00001
 &       &       &       &       &       & & \\
          &       & $\hat{b}$     & 3.294 &       &       &       &       &       & & \\
          \hline
    \multirow{2}[0]{*}{Lomax} & \multirow{2}[0]{*}{2} & $\hat{\beta}$  & 4637.821 & \multirow{2}[0]{*}{-796.59} & \multirow{2}[0]{*}{1597.2} & \multirow{2}[0]{*}{1601.9} & \multirow{2}[0]{*}{0.043} & \multirow{2}[0]{*}{0.085} & \multirow{2}[0]{*}{0.146} & \multirow{2}[0]{*}{0.964}\\
          &       & $\hat{\sigma}$ & 1.293 &       &       &       &       &       &  &\\
          \hline
    \multirow{2}[0]{*}{Log-gamma} & \multirow{2}[0]{*}{2} & $\hat{\alpha}$ & 40.569 & \multirow{2}[0]{*}{-790.41} & \multirow{2}[0]{*}{1584.8} & \multirow{2}[0]{*}{1589.6} & \multirow{2}[0]{*}{0.475} & \multirow{2}[0]{*}{0.308} & \multirow{2}[0]{*}{0.322} & \multirow{2}[0]{*}{0.975}\\
          &       & $\hat{\beta}$  & 4.938 &       &       &       &       &       & & \\
          \hline
    \multirow{2}[0]{*}{F\'rechet} & \multirow{2}[0]{*}{2} & $\hat{a}$     & 0.578 & \multirow{2}[0]{*}{-815.11} & \multirow{2}[0]{*}{1634.2} & \multirow{2}[0]{*}{1639} & \multirow{2}[0]{*}{0.002} & \multirow{2}[0]{*}{0.003} & \multirow{2}[0]{*}{0.005} & \multirow{2}[0]{*}{0.877} \\
          &       & $\hat{b}$     & 7684.803 &       &       &       &       &       & & \\
          \hline\hline
    \end{tabular*}%
  \label{tab:estimates - data I}%
\end{table}%

By considering the confidence levels of 95\% and 99\%,
Table \ref{tab:VaR-competing models} reports the empirical VaR as well as the model VaR based on the
fitted models. Percentage of variation of each model VaR with respect to the
empirical VaR, and ranking induced by the absolute value of this
measure, are also given to ease performance comparison.
At the most relevant 99\% VaR level,  GLMGA is closest to the empirical value.

\begin{table}[htbp]
  \centering
  \caption{VaR, difference (in percentage) with respect to the empirical VaR, ranking induced by the absolute difference for the competing models. }
   \begin{tabular*}{\hsize}{@{}@{\extracolsep{\fill}}ccccccc@{}}
   \hline\hline
    Model & 95\%  & Diff. \% & Ranking & 99\%  & Diff. \% & Ranking \\
    \hline
    Empirical & 57711.32 &   -    &   -    & 228895.4 &      - & - \\
    GlogM & 66850.13 &        15.84  & 3     & 572559.4 &      150.14  & 6 \\
    GB2   & 60193.16 &          4.30  & 1     & 536891.6 &      134.56  & 5 \\
    GLMGA & 45730.33 &      -20.76  & 4     & 296637.3 &        29.60  & 1 \\
    Lomax & 42388.83 &      -26.55  & 5     & 158614.7 &      -30.70  & 2 \\
    Log-gamma & 34359.6 &      -40.46  & 6     & 99571.8 &      -56.50  & 4 \\
    Fr\'echet & 51339.36 &      -11.04  & 2     & 108054.9 &      -52.79  & 3 \\
       \hline\hline
    \end{tabular*}%
  \label{tab:VaR-competing models}%
\end{table}%



\subsection{Application of the GLMGA regression to earthquake losses data set}
The regression methodology proposed in Section 4
is now applied to an earthquake loss data set of Chinese Mainland from Chinese Seismic Bureau (CSB) which contains risk information on
291 earthquake events
with magnitude greater than 4.0
from 1990 to 2015\footnote{
An earthquake resulting in one of several damage types such as casualties, economic losses, and damage to buildings
is defined to be one earthquakes event}.
The data set contains, among others,  the specific occurrence time, location, magnitude, seismic intensity,  and total economic damage of each earthquake event.

To quantify the utilization of earthquake risk information, we here study
the total economic losses, defined  as the direct economic losses associated with an earthquake impact as determined in the
weeks and sometimes months after the event.
The total economic damage is expressed in millions of Chinese Yuan (CNY) and is adjusted for inflation to reflect values in 2015 and is explained in terms of two covariates, the magnitude with values between 4.0 and 8.1, and seismic
intensity with categories 4, 5, 6, 7, 8, 9, and 11.
Table \ref{tab:earthquakes datset} reports the  major earthquake disasters in China since 1990. In particular, the 2008 earthquake in Sichuan is the most damaging earthquake. It took away about 69,227 lives and caused 3,757 billion Chinese Yuan (CNY) direct total economic damage.

From a preliminary analysis it follows that, for the earthquake events with
positive total economic damage, the distribution of this variable is right skewed with
long tails.
In Table \ref{tab:descrip} we present for some time windows the total numbers of earthquakes occurring, the median, mean, standard deviation, skewness and kurtosis of the total economic damage.
The large standard deviation and skewness
act as a first indicator of heavy tailedness.

\begin{table}[htbp]
  \centering
  \caption{Major Chinese earthquakes since 1990.}
      \begin{tabular*}{\hsize}{@{}@{\extracolsep{\fill}}cccccccc@{}}
    \hline
    \hline
    \tabincell{c}{Time of\\Occurrence} & Location & Magnitude & \tabincell{c}{Seismic\\Intensity} & \tabincell{c}{Deaths} & \tabincell{c}{Injuries} & \tabincell{c}{Economic\\Losses (original)} &\tabincell{c}{Economic\\Losses (adjusted)} \\
    \hline
    2013/08/31&Yunnan &5.9&8&3&63&1967&2834\\
    2012/06/30 & Xinjiang & 6.6   & 8     & 0     & 52    & 1990&3762
 \\
    2013/11/23 & Jilin & 5.5   & 7     & 0     & 25    & 2023& 2915
\\
    2005/11/26 & Jiangxi & 5.7   & 7     & 13    & 775   & 2038 &10827
\\
     2009/07/09 & Yunnan & 6.0    & 8     & 1     & 372   & 2154 &8196
\\
     2014/12/06 & Yunnan & 5.9   & 8     & 1     & 22    & 2377& 2771
\\
   2011/03/10 & Yunnan & 5.8   & 7     & 25    & 314   & 2385 &6146
\\
    1996/02/03 & Yunnan & 7.0     & 6    & 309   & 17057 & 2500 &46166
\\
    2013/08/12 & Tibet & 6.1   & 8     & 0     & 87    & 2707&3901
 \\
   2014/11/22 & Sichuan & 6.3   & 8     & 5     & 78    & 4232& 4934
\\
    2008/08/30 & Sichuan & 6.1   & 8     & 41    & 1010  & 4462& 18846
\\
    2012/09/07 & Yunnan & 5.7   & 8     & 81    & 834   & 4771&9018
 \\
 2014/10/07 & Yunnan & 6.6   & 8     & 1     & 331   & 5110 &5958
\\
2015/07/03 & Xinjiang & 6.5   & 8     & 3     & 260   & 5430 &5430
\\
    2015/04/25 & Tibet & 8.1   & 9     & 27    & 860   & 10302&10302
 \\
  2014/08/03 & Yunnan & 6.5   & 9     & 617   & 3143  & 19849& 23144
\\
 2010/04/14 & Qinghai & 7.1   & 9     & 2698  & 11000 & 22847& 74258
\\
2013/07/22 & Gansu & 6.6   & 8     & 95    & 2414  & 24416 &35182
\\
2013/04/20 & Sichuan & 7.0     & 9     & 196   & 13019 & 66514& 95841
\\
   2008/05/12 & Sichuan & 8.0    & 11    & 69227 & 375783 & 845110& 3569560
\\
    \hline
    \hline
    \end{tabular*}%
  \label{tab:earthquakes datset}%
\end{table}%

\begin{table}[htbp]
  \centering
  \caption{Distribution of the total economic damage as a function of time.}
  \begin{tabular*}{\hsize}{@{}@{\extracolsep{\fill}}cccccccc@{}}
    \hline\hline
    Year  & \tabincell{c}{Total Numbers \\of Earthquakes} & \multicolumn{1}{l}{Mean} & \tabincell{c}{Standard\\ Deviation} & \multicolumn{1}{l}{Median} & \multicolumn{1}{l}{Kurtosis} & \multicolumn{1}{l}{Skewness} \\
    \hline
     (1990,1995] & 63    & 1456  & 2764  & 328   & 18.39 & 3.58 \\
          (1995,2000] & 61    & 2485  & 7063  & 552   & 27.63 & 4.79 \\
          (2000,2005] & 60    & 1057  & 2027  & 294   & 13.98 & 3.22 \\
          (2005,2010] & 46    & 80432 & 525996 & 279   & 43.98 & 6.55 \\
           (2010,2015] & 61    & 3838  & 13203 & 386   & 40.58 & 5.95 \\
           Total & 291   & 14573 & 209282 & 340   & 288.09 & 16.93 \\
             \hline \hline
    \end{tabular*}%
  \label{tab:descrip}%
\end{table}%

\begin{figure}[htbp]
	\centering
	\includegraphics[scale=0.7]{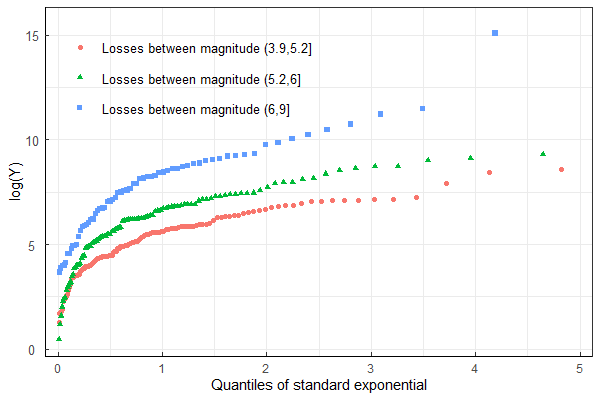}
	\caption{Pareto QQ-plots of economic losses for different intervals of magnitude levels.}
	\label{fig:Pareto QQ:loss}
\end{figure}


%

\newpage
\newpage

%
We now illustrate the GLMGA regression model from Section 4  modelling the total economic damage as a function of earthquake magnitude and intensity.
The estimation results of three GLMGA regression models are summarized
in Table \ref{tab:GLMGA} when the two covariates (magnitude and seismic intensity) including a constant term are introduced in  $\sigma_i$ and $b_i$, whether in only one or in both parameters.
As we would expect, the magnitude and seismic intensity are significant determinants of total economic damage in the three models.
In case of Model III, the extreme value index $2 \sigma$ runs between 0.65 at magnitude 4.0 and 0.96 at magitude 8.0 so that the (theoretical) second moments do not even exist.

Table \ref{tab:GLMGA} also reports the log-likelihood value, AIC  and BIC value of the proposed model.
We see that the $\text{GLMGA}(\sigma_i,a,b_i)$ model fits best  with the lowest AIC and BIC values. This model expresses that the magnitude does influence the tail heaviness parameter $\sigma_i$, while  the seismic intensity should be introduced in $b_i$. The Pareto QQ-plots of economic losses for different magnitude intervals in Figure \ref{fig:Pareto QQ:loss}  do indeed indicate an increase in extreme value index $2\sigma$ with increasing magnitude as the slopes of these plots at the largest loss levels  provide a graphical inspection of the extreme value index, see \cite{beirlant2004}.

\begin{table}[htbp]
  \centering
  \caption{Estimates of the GLMGA regression on the earthquake data set.}
  \begin{tabular*}{\hsize}{@{}@{\extracolsep{\fill}}lrrrrrrr@{}}
  \hline\hline
  \multicolumn{2}{c}{\multirow{3}[0]{*}{Parameters}}   &  \multicolumn{2}{c}{Model I} &  \multicolumn{2}{c}{Model II} & \multicolumn{2}{c}{Model III} \\
   \multicolumn{2}{c}{} & \multicolumn{2}{c}{$\text{GLMGA}(\sigma_i,a,b)$} & \multicolumn{2}{c}{$\text{GLMGA}(\sigma,a,b_i)$} & \multicolumn{2}{c}{$\text{GLMGA}(\sigma_i,a,b_i)$} \\
         \cmidrule(lr){3-4} \cmidrule(lr){5-6} \cmidrule(lr){7-8}
     \multicolumn{2}{c}{} & Estimate  & Std. error   &Estimate  & Std. error      & Estimate  & Std. error   \\
    \hline
    constant& $\beta_0$ & -2.54 & 0.14  &-0.97     & 0.11     & -1.52 & 0.17 \\
    magnitude& $\beta_1$& 0.10   & 0.02  & -     & -     & 0.10   & 0.02 \\
    intensity  &$\beta_2$ & 0.15  & 0.03  & -     & -     & -     & - \\
    constant &$\alpha_0$ & -17.27     & 2.16     & 5.11  & 1.93  & 2.15  & 2.01 \\
    magnitude &$\alpha_1$   & -     & -     & -1.18 & 0.42  & -2.91 & 0.39 \\
    intensity &$\alpha_2$ & -     & -     & -2.50  & 0.36  & -     & - \\
   parameter&$\eta$ & -1.21 & 0.18  & -1.27 & 0.18  & -1.15 & 0.17 \\
   \hline
    \multicolumn{2}{c}{Log-likehood}     & \multicolumn{2}{c}{-2235.3} & \multicolumn{2}{c}{-2236.7} & \multicolumn{2}{c}{\bf -2228.2} \\
     \multicolumn{2}{c}{AIC}     & \multicolumn{2}{c}{4480.7} & \multicolumn{2}{c}{4483.3} & \multicolumn{2}{c}{\bf 4466.5} \\
     \multicolumn{2}{c}{BIC}     & \multicolumn{2}{c}{4499.0} & \multicolumn{2}{c}{4501.7} & \multicolumn{2}{c}{\bf 4484.8} \\
    \hline\hline
    \end{tabular*}%
  \label{tab:GLMGA}%
\end{table}%


\begin{figure}[htbp]
	\centering
	\includegraphics[scale=0.6]{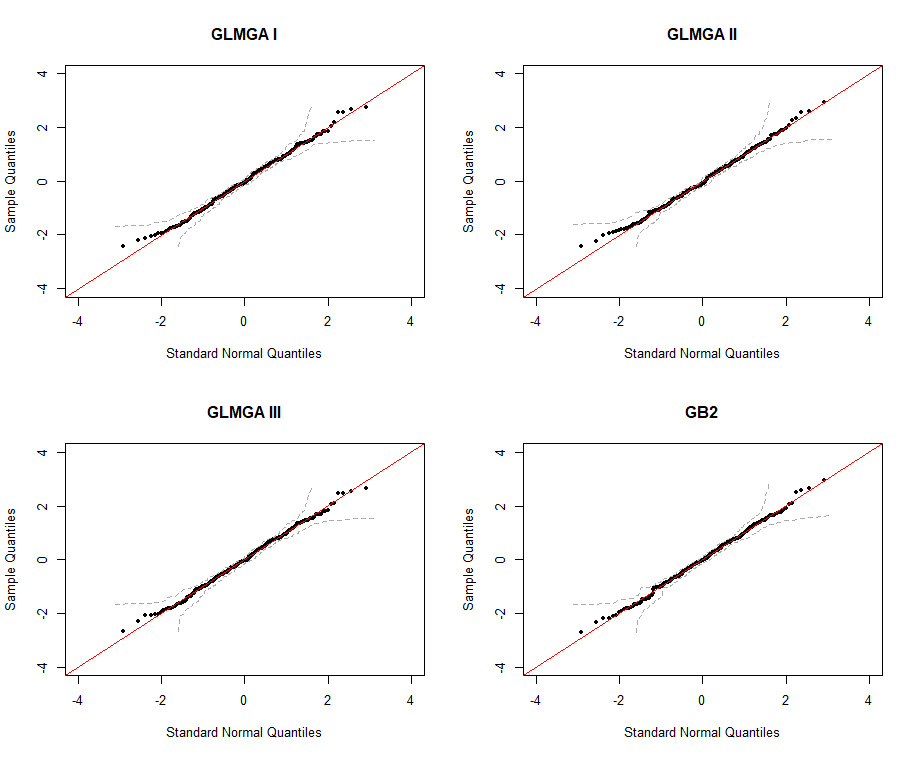} 
	\caption{Normal QQ-plots of quantile residuals $r_i$ from three GLMGA regression models considered in Table \ref{tab:regression setting}, together with the best GB2 regression model. }
	\label{fig:qqGLMGA}
\end{figure}

We  also compare the proposed GLMGA regression models with the generalized log-Moyal regression model discussed in \cite{bhati2018generalized},
the exponentiated Fr\' echet regression model recently discussed in \cite{gunduz2016exponentiated}, the
GB2 regression model that is widely used in non-life insurance rate-making \citep{shi2015private, frees2008hierarchical},
the Burr regression \citep{beirlant1998burr},  the
exponential-inverse Gaussian regression \citep{frangos2004modelling},
the gamma-generalized inverse Gaussian regression \citep{gomez2013gamma},
the lognormal regression model discussed in \cite{stasinopoulos2007generalized}, next to traditional generalized linear models (gamma regression, inverse Gaussian regression)
respectively. The regression models above are given in Table \ref{tab:regression setting} using the parameter notation from the original papers. Specifically several regression models were fitted on the GB2 model and the proposed model fitting the GB2 parameter $\mu$  as a function of magnitude and intensity, which corresponds to $(2b)^{1/\sigma}$ in the GLMGA parametrization, came out best using the different criteria.

To demonstrate the goodness of fit of the GLMGA regression,
we provide in Figure \ref{fig:qqGLMGA} the QQ-plots of the randomized quantile
residuals $r_{i}=\Phi^{-1}\left[F(y_{i})\right]$.
The dashed lines indicate
95\% confidence intervals.
The match between the
theoretical and empirical quantiles suggests the favorable fit
of the GLMGA distribution with correlation coefficient 0.998 for the GLMGA III model and 0.963 for the best fitting GB2 model.

Table \ref{tab:compare models} presents a model comparison in terms of goodness-of-fit.
Model $\text{GLMGA}(\sigma_i,a,b_i)$  has the  highest loglikelihood value.
Rankings
induced by AIC and BIC value put the $\text{GLMGA}(\sigma_i,a,b_i)$  and Lomax models at the top.
The second best
is the GB2 model,  followed by the $\text{GLMGA}(\sigma_i,a,b)$ and $\text{GLMGA}(\sigma,a,b_i)$ models.
\begin{table}[htbp]
  \centering
  \caption{Earthquake economic losses: log-likelihood, AIC, and BIC for the competing models, along with rankings induced by these
criteria.}
    \begin{tabular*}{\hsize}{@{}@{\extracolsep{\fill}}lcrrrrr@{}}
    \hline\hline
    Regression models &  \#Par.&Log-like. & AIC   &Ranking& BIC&Ranking \\
    \hline
    $\text{GLMGA}(\sigma_i,a,b_i)$ &5& \bf -2228.2 & \bf 4466.5 &\bf 1& \bf 4484.8  &\bf 1\\
    GB2   &6& -2228.9 & 4469.8 &2& 4491.8  &2\\
    $\text{GLMGA}(\sigma_i,a,b)$ &4& -2235.3 & 4480.7 &5& 4499.0  &3\\
    Gamma-generalized inverse Gaussian &6& -2229.9 & 4473.9 &4& 4499.6  &4\\
    Lognormal &4& -2238.6 & 4485.3 & 7&4500.0  &5\\
    $\text{GLMGA}(\sigma,a,b_i)$ &5& -2236.7 & 4483.3 &5& 4501.7  &6\\
    Burr  &5& -2236.7 & 4483.4 & 6&4501.8  &7\\
    Exponentiated Fr\'echet&5& -2240.1 & 4490.2 & 8&4508.5  &8\\
    Gamma&4& -2256.6 &4521.2 & 9&4535.9  &9\\
    Weibull &4& -2277.2 & 4562.4 & 10&4577.1  &10\\
    Exponential-inverse Gaussian &4&-2280.0&4568.1&11&4582.8&11\\
    Generalized Pareto &4& -2288.2&4584.5&12&4599.2&12\\
    GlogM &4& -2316.4 & 4640.8 & 13&4655.5  &13\\
    Inverse Gaussian &4&-2363.4&4734.7&14&4749.4&14\\
    \hline\hline
    \end{tabular*}%
  \label{tab:compare models}%
\end{table}%

\newpage
\section{Concluding remarks}\label{section:conclusion}
In this paper we proposed a new heavy tailed model which appears to be a good candidate to model heavy tailed insurance data, not the least in regression analysis. It provides a useful submodel from the popular GB2 model.
Since an insurer's risk is often measured  by several  variables jointly,  a multivariate extension of the generalized log-Moyal gamma distribution through a copula construction should be the subject of future research.

\section*{Acknowledgement}
Zhengxiao Li and Shengwang Meng acknowledge the
financial support from National Natural Science Fund of China (Grant No. 71901064), National Social Science Fund of China (Grant No. 16ZDA052)
and MOE National Key Research Bases for Humanities and Social Sciences (Grant No.
16JJD910001) .

\section*{Declaration of interest }
We declare that there is no potential conflict of interest in the paper.

\bibliographystyle{biom}

\newpage
\section*{Appendix}
\appendix
  \renewcommand{\appendixname}{Appendix~\Alph{section}}
\section{Earthquake economic losses: the competing regression models}
\begin{table}[htbp]
  \centering
  \caption{Probability density distribution functions and regression components used.}
    \begin{tabular*}{\hsize}{@{}@{\extracolsep{\fill}}lll@{}}
    \hline\hline
    Distribution  & \tabincell{c}{ Density function} & \tabincell{l}{Regression \\component} \\
    \hline
    lognormal &$\frac{1}{\sqrt{2\pi}y_i}\exp\left[-\frac{(\log y_i-\mu_i)^2}{2\sigma^2}\right]$&$\mu_i=\bm{x}_{i}^{T}\bm{\beta}$\\[15pt]
    Burr    &$\frac{\lambda\theta^{\lambda}\tau_i{y_i}^{\tau_i-1}}{\left(\theta+{y_i}^{\tau_i}\right)^{\lambda+1}}$&$\tau_i=\exp(\bm{x}_{i}^{T}\bm{\beta})$\\[15pt]
    Generalized log-Moyal &$\frac{\sqrt{\theta}}{\sqrt{2\pi }\sigma_i }\ {{\left( \frac{1}{y_{i}} \right)}^{\frac{1}{2\sigma_i }+1}}{{\exp}{\left[-\frac{\theta}{2}{{\left( \frac{1}{y_{i}} \right)}^{1/\sigma_i}}\right]}}$&$\sigma_i=\exp(\bm{x}_{i}^{T}\bm{\beta})$\\[15pt]
 Exponentiated Fr\' echet& $\alpha\lambda_i\tau\left\{
1-\exp\left[-\left(\frac{\tau}{{y_i}^\lambda_i}\right)\right]
\right\}^{\alpha-1}y^{-(1+\lambda_i)}\exp\left[-\left(\frac{\tau}{{y_i}^\lambda_i}\right)\right]$  &  $\lambda_i=\exp(\bm{x}_{i}^{T}\bm{\beta})$\\[15pt]
\tabincell{c}{Gamma-generalized\\inverse Gaussian}&$\frac{(y_i t_i\mu)^a}{y_i \Gamma(a)}\left(\frac{\psi}{\psi+2\mu^2t_i y_i}\right)^{\frac{a+\lambda}{2}}\frac{K_{a+\lambda}(\psi/\mu_{1}^{*})}{K_{\lambda}(\psi/\mu)}$
& $t_i=\exp(\bm{x}_{i}^{T}\bm{\beta})$\\[15pt]
\tabincell{c}{Exponential-inverse\\Gaussian}&
$\frac{\delta}{t_i}\frac{\exp\left[-\delta\phi(y_i,\bm{\beta},\delta)-\delta^2\right]}{\phi(y_i,\beta,\delta)^3}\left[
\delta\phi(y_i,\bm{\beta},\delta)+1
\right]$
&
$t_i=\exp(\bm{x}_{i}^{T}\bm{\beta})$\\[15pt]
GB2&$\frac{|\sigma|}{y_iB(\nu,\tau)}\frac{(y_i/\mu_i)^{p v}}{\left[1+\left(y_i/\mu_i\right)^p\right]^{\tau+\nu}}$&$\mu_i=\exp(\bm{x}_{i}^{T}\bm{\beta})$\\
Gamma &$\frac{1}{\beta_i^{\alpha}}y_{i}^{\alpha-1}\exp\left(-\frac{y_i}{\beta_i}\right)$
&$\beta_i=\exp(\bm{x}_{i}^{T}\bm{\beta})$\\[15pt]
Inverse Gaussian&$\left(\frac{\lambda}{2\pi y_i^{3}}\right)^{1/2}\exp\left[-\frac{\lambda(y_i-\mu_i)^2}{2\mu_i^2y_i}\right]$
&$\mu_i=\exp(\bm{x}_{i}^{T}\bm{\beta})$\\[15pt]
Generalized Pareto &$\frac{1}{\sigma}\left[1+\frac{\xi_i(y-\mu)}{\sigma}\right]^{\left(-\frac{1}{\xi_i}-1\right)}$&\tabincell{l}{$\xi_i=\bm{x}_{i}^{T}\bm{\beta}$ \\ $\mu=0$}\\[15pt]
    \hline\hline
    \end{tabular*}%
      \vspace{1ex}
     {\raggedright \small{
     Note: In the gamma-generalized inverse Gaussian regression model, $\mu_{1}^{*}=\mu\sqrt{\frac{\psi}{\Psi+2\mu^2t_i y_i}}$, $K_{m}(\cdot)$ is the
modified bessel function of the third kind with order $m$.
In the exponential-inverse Gaussian regression model,
$\phi(y_i,\bm{\beta},\delta)=(\delta^2+2y_{i}/t_{i})^{1/2}$.
\par}}
  \label{tab:regression setting}%
\end{table}

\newpage
\renewcommand{\appendixname}{Appendix~\Alph{section}}
\section{Proofs}

{\bf Proof of Proposition 2.1.}


Note that when $Y\sim \text{GLMGA}(\sigma,a,b)$,
we derive the marginal distribution $F(y)$ as follows.
By letting $z=y^{-\frac{1}{\sigma}}$ and $v=\frac{z}{z+2b}$,
we have
\begin{align}
F(y)&=\int_{0}^{y}\frac{b^a}{\sqrt{2}\sigma B\left(a,\frac{1}{2}\right)}\frac{y^{-\left(\frac{1}{2\sigma}+1\right)}}{\left(\frac{1}{2}y^{-\frac{1}{\sigma}}+b\right)^{a + \frac{1}{2}}}dy \nonumber \\
&=1-\int_{0}^{y^{-\frac{1}{\sigma}}}\frac{1}{zB(a,\frac{1}{2})}\frac{z^{\frac{1}{2}}(2b)^a}{(z+2b)^{a+\frac{1}{2}}}dz \nonumber \\
&=1-\int_{0}^{\frac{z}{z+2b}}\frac{1}{B(a,\frac{1}{2})}v^{-\frac{1}{2}}(1-v)^{a-1}dv \nonumber \\
&=1-I_{\frac{1}{2},a}\left[\frac{z}{z+2b}\right].
\label{cdf:LMGA}
\end{align}
The $u^{th}$ quantile of $F^{-1}(u)$ of $\text{GLMGA}(\sigma,a,b)$ can be obtained by inverting the cdf (\ref{cdf:LMGA}).


\vspace{0.5cm}\noindent
{\bf Proof of \eqref{lim_density}.}
Let $b=\frac{a}{2}\phi^{\frac{1}{\sigma}}$ in $\text{GLMGA}(\sigma, a, b)$. Substituting this  in (\ref{pdf:LMGA}) yields
\begin{align*}
f(y;\sigma, a, b)
&=\frac{b^a}{\sqrt{2}\sigma B\left(a,\frac{1}{2}\right)}\frac{y^{-\left(\frac{1}{2\sigma}+1\right)}}{\left(\frac{1}{2}y^{-\frac{1}{\sigma}}+b\right)^{a + \frac{1}{2}}}\\
&=f(y;\sigma,a,\phi)\\
&=\frac{y^{-\frac{1}{2\sigma}-1}}{\sigma a^{\frac{1}{2}}\phi^{\frac{1}{2\sigma}}B(\frac{1}{2},a)}\left[\frac{1}{1+\frac{y^{-1/\sigma}}{2b}}
\right]^{a+\frac{1}{2}}\\
&=\frac{y^{-\frac{1}{2\sigma}-1}}{\sigma\sqrt{\pi}\phi^{\frac{1}{2\sigma}}}\left[\frac{1}{1+\frac{y^{-1/\sigma}}{2b}}
\right]^{a+\frac{1}{2}}\frac{\Gamma(a+\frac{1}{2})}{\Gamma(a)a^{\frac{1}{2}}}.
\end{align*}
For large values of $a$, the gamma function can be approximated by Stirling's formula thus
\[
\lim_{a\to\infty}\frac{\Gamma(a+\frac{1}{2})}{\Gamma(a)a^\frac{1}{2}}=
\frac{e^{-\frac{1}{2}-a}(a+\frac{1}{2})^{a+\frac{1}{2}-\frac{1}{2}\sqrt{2\pi}}}{e^{-a}a^{a-\frac{1}{2}\sqrt{2\pi}}a^\frac{1}{2}}=1.
\]
Similarly,
\[
\lim_{a\to\infty}\left[\frac{1}{1+\frac{y^{-1/\sigma}}{2b}}
\right]^{a+\frac{1}{2}}
=\lim_{a\to\infty}\left[\frac{a}{a+(\phi y)^{-1/\sigma}}
\right]^{a+\frac{1}{2}}
=\lim_{a\to\infty}\exp\left[-\left(\phi y\right)^{-\frac{1}{\sigma}}\right].
\]
For $a\to\infty$, we have
\[
\lim_{a\to\infty}f(y)=
\frac{1}{\sigma\sqrt{\pi}\phi^{\frac{1}{2\sigma}}}y^{-\frac{1}{2\sigma}-1}\exp\left[-\left(\phi y\right)^{-\frac{1}{\sigma}}\right],
\]
which is the density of  generalized inverse gamma distribution with shape parameters $\frac{1}{2}$ and  $\frac{1}{\sigma}$, and scale parameter $\phi$. \\
For $\sigma=\frac{1}{2}$ and $a\to\infty$, we have
\[
\lim_{a\to\infty}f(y)=\frac{2}{\phi\sqrt{\pi}y^2}\exp\left(-\frac{1}{\phi^2y^2}\right),
\]
which is the pdf of inverse half-normal distribution.

\vspace{0.4cm}\noindent
{\bf Proof of \eqref{incomplete_moment_bottom} and \eqref{incomplete_moment_top}.}
By letting $z=y^{-\frac{1}{\sigma}}$ and $v=\frac{z}{z+2b}$, we have
\begin{align*}
\int_{0}^{u}y^kf(y)dy&=\int_{0}^{u}y^h\frac{b^a}{\sqrt{2}\sigma B\left(a,\frac{1}{2}\right)}\frac{y^{-\left(\frac{1}{2\sigma}+1\right)}}{\left(\frac{1}{2}y^{-\frac{1}{\sigma}}+b\right)^{a + \frac{1}{2}}}dy\\
&=\int_{u^{-\frac{1}{\sigma}}}^{+\infty}z^{-h\sigma-\frac{1}{2}}\frac{1}{B(a,\frac{1}{2})}\frac{(2b)^a}{(z+2b)^{a+\frac{1}{2}}}dz \nonumber \\
&=(2b)^{-h\sigma}\frac{B(\frac{1}{2}-h\sigma, a+h\sigma)}{B(a,\frac{1}{2})}\left[1-\int_{0}^{\frac{u^{-1/\sigma}}{u^{-1/\sigma}+2b}}
\frac{v^{\frac{1}{2}-h\sigma-1}(1-v)^{a+h\sigma-1}}{B(\frac{1}{2}-h\sigma, a+h\sigma)}
dv\right] \nonumber \\
&=\mathbb{E}(Y^h)\left\{1-I_{\frac{1}{2}-h\sigma,a+h\sigma}\left[\frac{u^{-1/\sigma}}{u^{-1/\sigma}+2b}\right]\right\}.
\end{align*}
Similarly,
\[
\int_{u}^{+\infty}y^kf(y)dy=\mathbb{E}(Y^h)I_{\frac{1}{2}-h\sigma,a+h\sigma}\left[\frac{u^{-1/\sigma}}{u^{-1/\sigma}+2b}\right].
\]

\newpage
\renewcommand{\appendixname}{Appendix~\Alph{section}}
\section{Extra Figures}

\begin{figure}[htbp]
	\centering
	\includegraphics[scale=0.8]{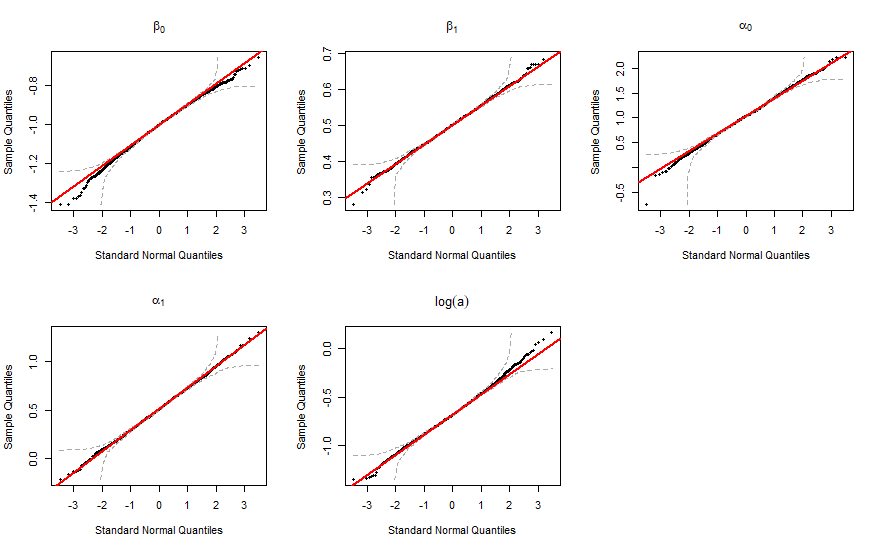} 
	\caption{Normal QQ-plots of ML parameter estimators  from GLMGA regression when sample size $n$ = 200 and $(\beta_0,\beta_1,\alpha_0,\alpha_1,a) = (-1,0.5,1,0.5,0.5)$. }
	\label{fig:size200-a=0.5}
\end{figure}

\begin{figure}[htbp]
	\centering
	\includegraphics[scale=0.8]{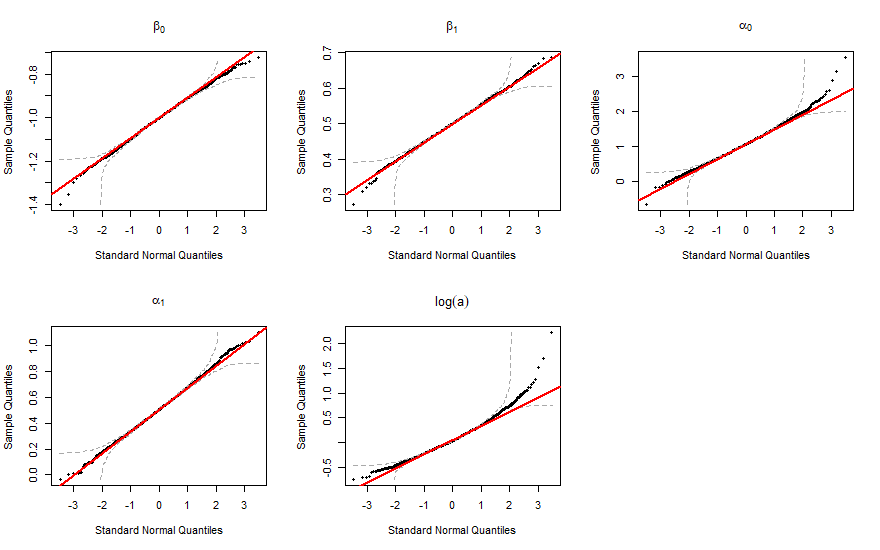} 
	\caption{Normal QQ-plots of ML parameter estimators from GLMGA regression when sample size $n$ = 200 and $(\beta_0,\beta_1,\alpha_0,\alpha_1,a) = (-1,0.5,1,0.5,1)$.  }
	\label{fig:size200-a=1}
\end{figure}

\begin{figure}[htbp]
	\centering
	\includegraphics[scale=0.8]{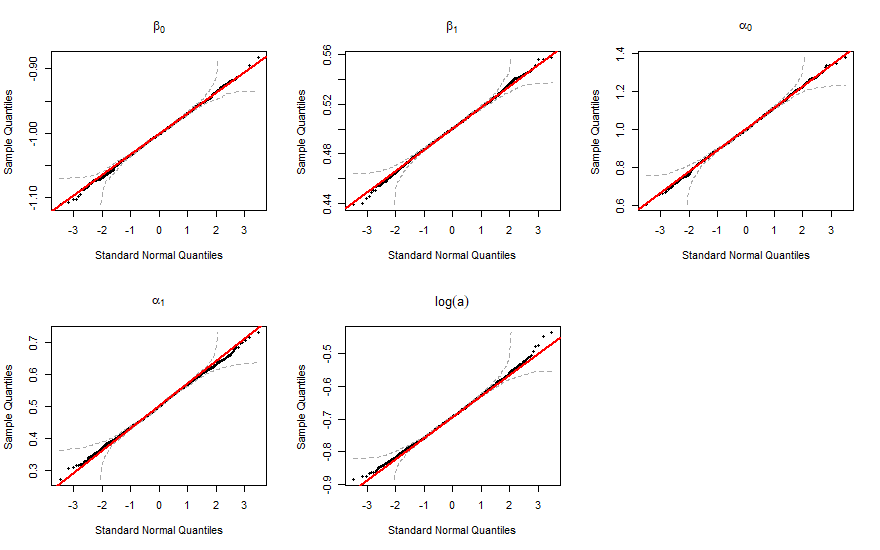} 
	\caption{Normal QQ-plots of ML parameter estimators from GLMGA regression when sample size $n$ = 2000 and $(\beta_0,\beta_1,\alpha_0,\alpha_1,a) = (-1,0.5,1,0.5,0.5)$.  }
	\label{fig:size1000-a=0.5}
\end{figure}

\begin{figure}[htbp]
	\centering
	\includegraphics[scale=0.8]{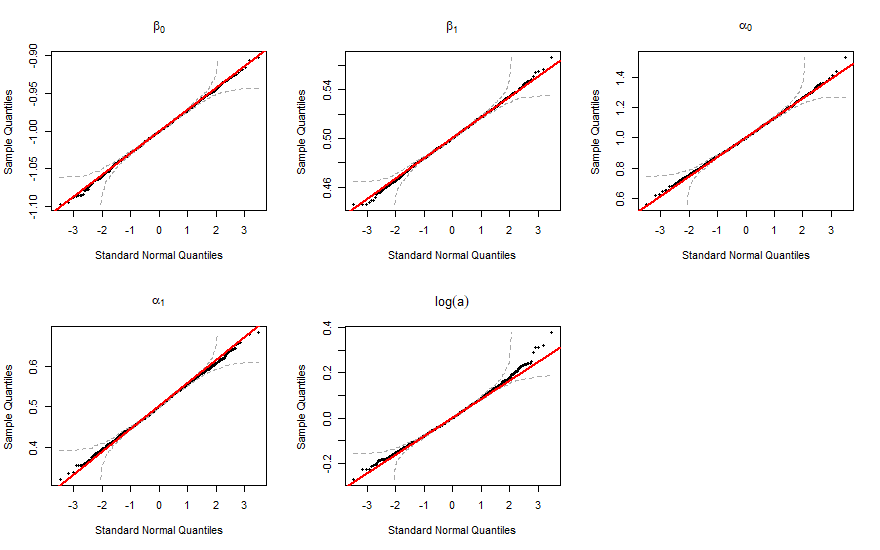} 
	\caption{Normal QQ-plots of ML parameter estimators from GLMGA regression when sample size $n$ = 2000 and $(\beta_0,\beta_1,\alpha_0,\alpha_1,a) = (-1,0.5,1,0.5,1)$.  }
	\label{fig:size1000-a=1}
\end{figure}

\end{document}